\documentclass[12pt]{article}

\usepackage{amssymb}
\usepackage{amsmath}
\usepackage{graphicx}
\usepackage{color}

\def\({\left(}
\def\){\right)}
\def\wt#1{\widetilde{#1}}
\def\wh#1{\widehat{#1}}
\def\tensor{\otimes}

\def\W{\ell_{0}}
\def\NABLA#1{\nabla^{(#1)}}
\def\LAP#1{\text{Lap}^{(#1)}}
\def\DIV#1{\text{Div}^{(#1)}}
\def\EIN#1{\text{Ein}^{(#1)}}
\def\TR#1{\text{Tr}^{(#1)}}
\def\RIC#1{\text{Ric}^{(#1)}}
\def\MAN#1{\mathcal{M}^{(#1)}}
\def\TRREV#1{\mu^{(#1)}}

\def\CVRR#1{ \mathcal{R}^{g_{#1}} }
\def\HMAX{\mathcal{H}}
\def\taumax{\tau_{\text{max}}}
\def\AR{\mathcal{A}(\taumax)}
\def\RDO{\dot{R}(0)} 
\def\ZDO{\dot{Z}(0)} 
\def\thDO{\dot{\theta}(0)} 
\newcommand{\CF}{{ \mathcal C}}
\newcommand{\UU}{{ \mathcal U}}
\newcommand{\UUP}{{ \mathcal P}_{\mathcal U}}
\newcommand{\df}{\frac{}{}}
\def\LASERSPACECURVES#1#2{
	\begin{figure}[!ht]
		\centering
		\includegraphics[width=#1\textwidth]{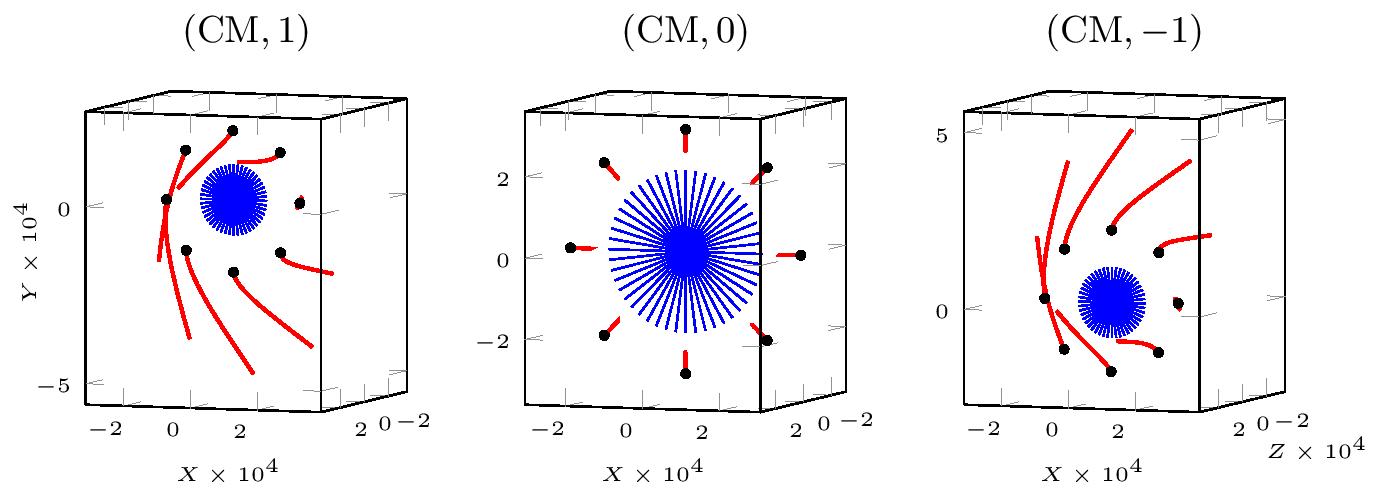}
		\caption{#2}
		\label{fig:laserspacecurves}
	\end{figure} 
	}
\def\HMAXRRZP#1#2{
	\begin{figure}[!ht]
		\centering
		\includegraphics[width=#1\textwidth]{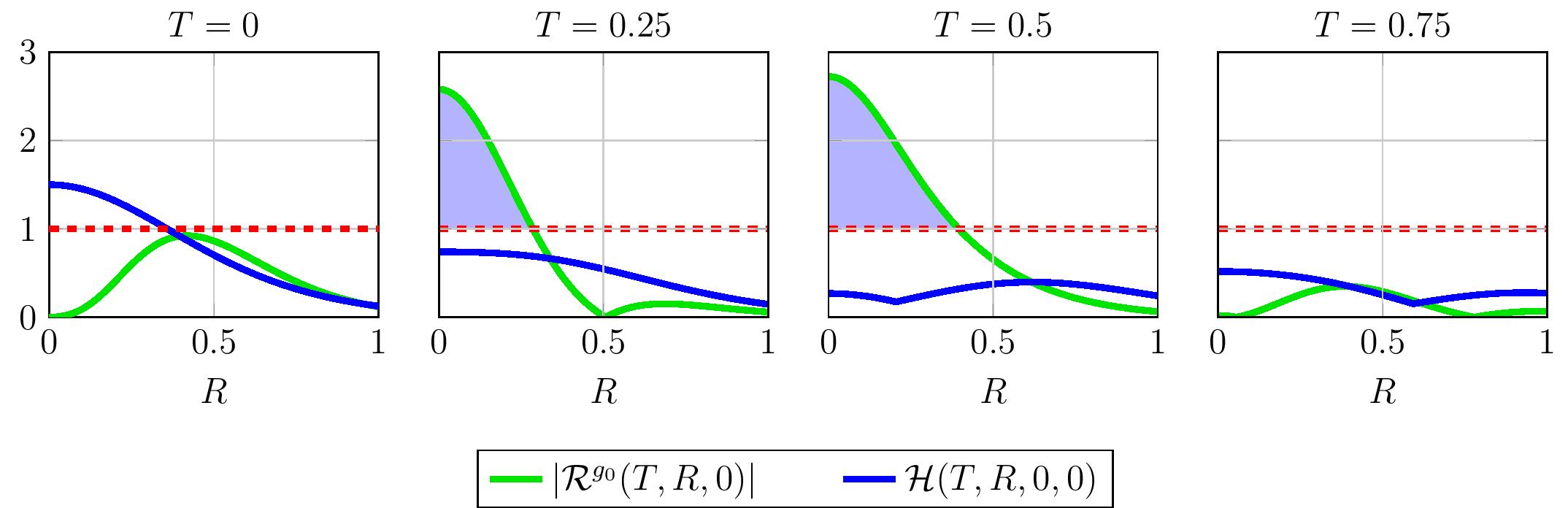}
		\caption{ #2 }
		\label{fig:hmaxrrzp}
	\end{figure}
	}
\def\RRZP#1#2{
	\begin{figure}[!ht]
		\centering
		\includegraphics[width=#1\textwidth]{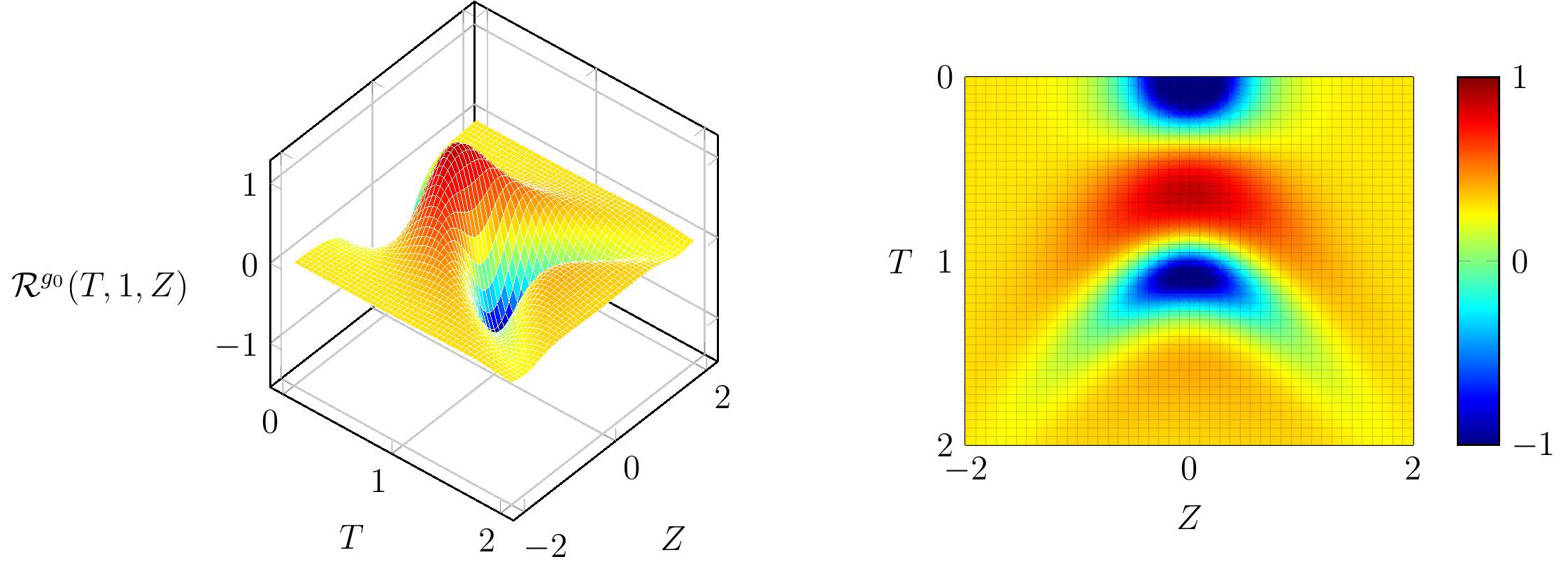}
		\caption{ #2 }
		\label{fig:rrzp}
	\end{figure}
	}
\def\ZPZLAYERS#1#2{
	\begin{figure}[!ht]
		\centering
		\includegraphics[width=#1\textwidth]{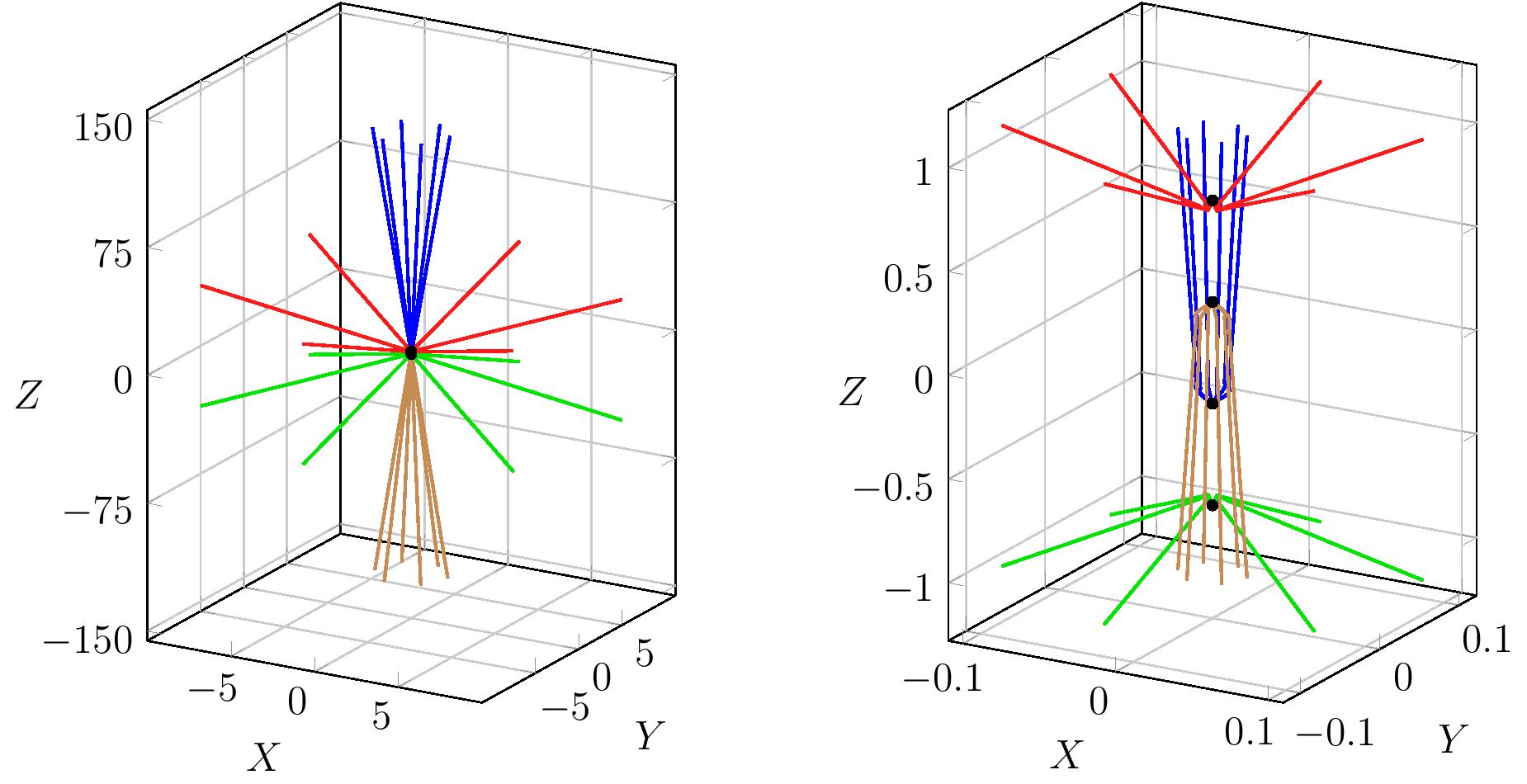}
		\caption{ #2 }
		\label{fig:zpzlayers}
	\end{figure}
	}
\def\ZPZLAYERSQ#1#2{
	\begin{figure}[!ht]
		\centering
		\includegraphics[width=#1\textwidth]{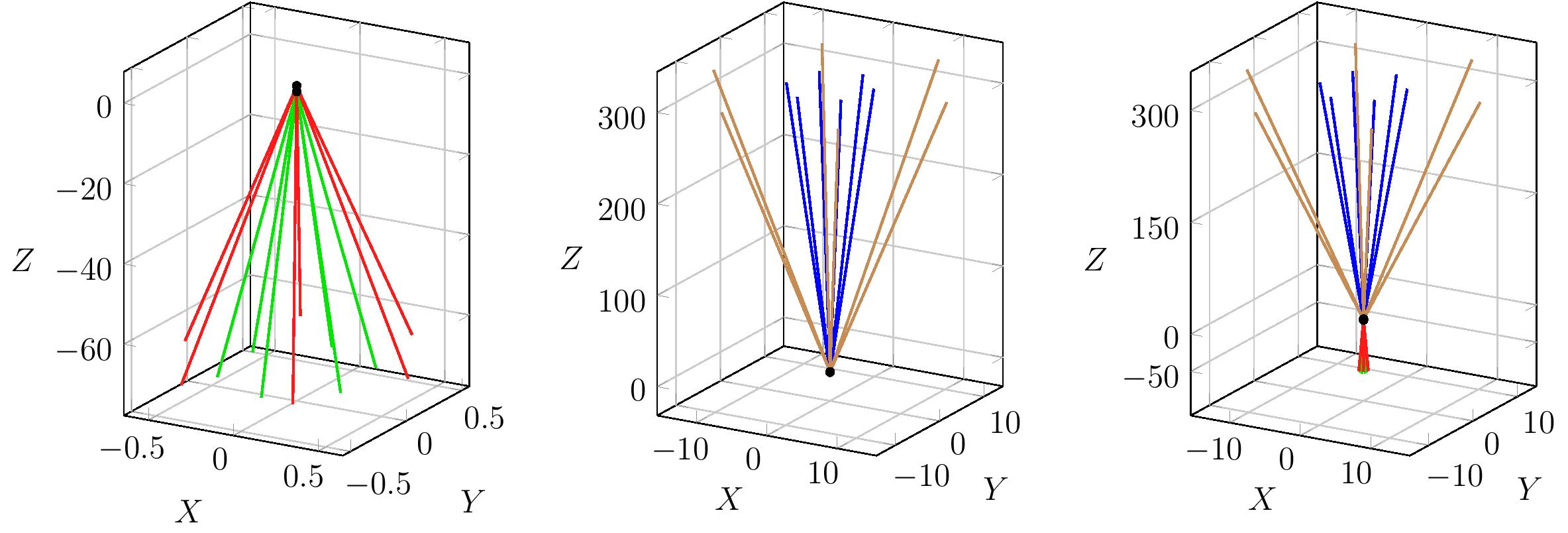}
		\caption{ #2 }
		\label{fig:zpzlayersq}
	\end{figure}
	}
	
\begin{document}

\title{Chirality in Gravitational and Electromagnetic Interactions with Matter}
\author{Robin W. Tucker${}^{1}$ and Timothy J. Walton${}^{2,}$\footnote{Corresponding author: t.walton@bolton.ac.uk} \\[0.2cm]
	\begin{tabular}{p{0.9\textwidth}} 
		\footnotesize ${}^{1}$ Department of Physics, University of Lancaster and Cockcroft Institute, Daresbury Laboratory, Warrington, UK \\[-0.6cm]
		\footnotesize${}^{2}$ Department of Mathematics, University of Bolton, Deane Campus, Bolton, UK \\
	\end{tabular}
 }

\maketitle

\begin{abstract}
	It has been suggested that single and double jets observed emanating from certain astrophysical objects may have a purely gravitational origin. We discuss new classes of pulsed gravitational wave solutions to the equation for perturbations of Ricci-flat spacetimes around Minkowski metrics, as models for the genesis of such phenomena. 
	We discuss how these solutions are motivated by the analytic structure of spatially compact finite energy pulse solutions of the source-free Maxwell equations generated from complex chiral eigen-modes of a chirality operator.
	Complex gravitational pulse  solutions  to the linearised source-free Einstein equations are classified in terms of their {\it chirality} and generate a family of non-stationary real spacetime metrics. Particular members of these families are used as backgrounds in analysing time-like solutions to the geodesic equation for test particles. They are found numerically to exhibit {\it both} single and double jet-like features with dimensionless aspect ratios suggesting that it may be profitable to include such backgrounds in simulations of astrophysical jet dynamics from rotating accretion discs involving electromagnetic fields. 
\end{abstract}\quad \\

\section{Introduction}
Many astrophysical phenomena find an adequate explanation in the context of Newtonian gravitation and Einstein's description of gravitation (together with Maxwell's theory of electromagnetism and the use of time-like spacetime geodesics to model the histories of massive point test particles) is routinely used to analyse a vast range of phenomena where non-Newtonian effects are manifest. However, there remain a number of intriguing astrophysical phenomena suggesting that our current understanding is incomplete. These include the large scale dynamics of the observed Universe and a detailed dynamics of certain compact stellar objects interacting with their environment. \\

In this note we address the question of the dynamical origin of the extensive ``cosmic jets'' that have been observed emanating from a number of compact rotating sources. Such jets often contain radiating plasmas and are apparently the result of matter accreting on such sources in the presence of magnetic fields. One of the earliest models to explain these processes suggested that the gravitational fields of rotating black holes surrounded by a magnetised ``accretion disc'' could provide a viable mechanism \cite{blandford_z}. More recently, the significance of magneto-hydrodynamic processes in transferring angular momentum and energy into collimated jet structures has been recognised \cite{blandford,lynden,pringle}. Many of these models implicitly assume the existence of a magnetosphere in a {\it stationary} gravitational field and employ ``force-free electrodynamics'' in their development. To our knowledge, a dynamical model that fully accounts for all the observed aspects of astrophysical jets does not exist.\\

However in recent years there has been mounting evidence, both theoretical and numerical, suggesting that non-Newtonian gravitational fields may be relevant for their genesis. By the genesis of such phenomena we mean a mechanism that initiates the plasma collimation process whereby electrically charged matter arises from initial distributions of neutral matter in a background gravitational field. In \cite{mashhoonPRD,mashhoonPLA}, the authors carefully analyse the properties of a class of Ricci-flat cylindrically symmetric spacetimes possessing time-like and null geodesics that approach attractors confining massive particles to cylindrical spacetime structures. Additional studies \cite{mashhoonJets,mashhoonPec,mashoonTidal} of the asymptotic behaviour of test particles on time-like geodesics with large Newtonian speeds relative to a class of co-moving observers have given rise to the notion of cosmic jets associated with different types of gravitational collapse scenarios satisfying certain Einstein-Maxwell field systems. There has also been a recent approach based on certain approximations within a linearised gravitational framework involving ``gravito-magnetic fields'' generated by non-relativistic matter currents \cite{poirier}. All these investigations auger well for the construction of models for astrophysical jets that include non-Newtonian gravitational fields as well as electromagnetically induced plasma interactions.\\

Although {\it astrophysical jets} involve both gravitational and electromagnetic interactions with matter it is natural to explore the structure of electrically neutral test particle geodesics in non-stationary, anisotropic background metric spacetimes as a first approximation. We report here on the construction of particular exact solutions to the linearised Einstein vacuum equations which are then used to numerically calculate time-like geodesics in such backgrounds. The use of the linearised Einstein vacuum equations facilitates the construction of a family of complex solutions with definite {\it chirality} that are used to demonstrate the existence of real spacetime metrics exhibiting families of time-like geodesics possessing particular jet-like characteristics on space-like hyper-surfaces. Test particles on such time-like geodesics exhibit, in general, a well defined sense of ``handed-ness'' in space that we argue may offer a mechanism that initiates a uni-directional jet structure. In particular, we construct families of new non-stationary metrics having propagating pulse-like characteristics with bounded components in three-dimensional spatial domains. The derivation of this class of solutions is based on a methodology used to construct single- or few-cycle laser pulse solutions to the vacuum Maxwell solutions in Minkowski spacetime \cite{GTW_PIPAMON}.  To facilitate the construction of gravitational pulse-like solutions this methodology will be reviewed first.

\section{Electromagnetic Pulses in Vacua}
The derivation of  finite energy solutions of the source-free Maxwell equations has a long history. The relevance of such solutions to modern technology has become apparent  with the advent of the laser. Since a spacetime description of the electromagnetic field employs spacetime antisymmetric tensors the language of exterior differential forms is appropriate.  Then the  source-free vacuum Maxwell system for the electromagnetic 2-form $F$ is
\begin{eqnarray}\label{SFM}
	d\, F \,=\, 0, \quad	\delta\, F \,=\, 0 
\end{eqnarray}
in terms of the exterior derivative linear operator $d$, the co-derivative linear operator $\delta\,\equiv\,\star\, d\,\star$ and the linear Hodge star map $\star$. These operators satisfy $d\circ d=0$ and  $ \delta\circ \delta=0$ since in a Lorentzian spacetime $\star (\star {\cal A}) = -(-1)^{p(4-p)}\,{\cal A} $ for any $p-$form ${\cal A}$. A complex 2-form $\Pi$ is said to be closed and co-closed if it satisfies the relations $d\Pi =0$ and $\delta\Pi = 0$ respectively. It follows that such a 2-form is covariantly constant: $\nabla \Pi=0$  where $\nabla$ denotes the Levi-Civita covariant differential. Complex solutions of (\ref{SFM}) can be generated in terms of such a 2-form and a complex 0-form $\alpha$ by writing $F=d\,A$ where 
\begin{eqnarray*}
	A &=& \star\,d\(\alpha\Pi\)
\end{eqnarray*}
provided 
\begin{eqnarray}\label{BOX}
	\delta d \alpha \,=\, 0.
\end{eqnarray}
This follows since 
\begin{eqnarray*}
	d \star d A &=& d \delta d \alpha \wedge \Pi .   
\end{eqnarray*}
Equation (\ref{BOX}) has many solutions. In local coordinates $\{x^{\mu}\}\equiv\{t,x,y,z\}$ with the Minkowski spacetime metric
\begin{eqnarray*}
	g &=& -c_{0}^{2}\,d t \otimes d t + d x \otimes d x  + d y \otimes d y + d z \otimes d z,
\end{eqnarray*}
a particularly simple class of finite-energy solutions that can be generated in this way follows from the complex axi-symmetric scalar solution \cite{Britt,Synge,Ziol85,Ziol89}
\begin{eqnarray}\label{alp}
   \alpha(t,r,z) &=& \frac{\W^{2}}{r^{2} + ( \psi_{1} + i(z-c_{0}t))\,(  \psi_{2}-i( z+c_{0}t ))  }
\end{eqnarray}
where $r^{2}=x^{2}+y^{2}$ and $\W,\psi_{1},\psi_{2}$ are real constants. One may generate a complex six dimensional chiral eigen-basis of covariantly constant 2-forms $\Pi^{s,\,\chi}$ satisfying 
\begin{eqnarray*}
	{ \cal L  }_{\partial_{\theta}}  \, \Pi^{s,\,\chi} &=& \chi\,\Pi^{s,\,\chi }
\end{eqnarray*}
with $s \in \{\mathrm{CE}, \mathrm{CM}\}$, $\chi \in \{ {1}, 0, -1\}$ where $x=r\cos(\theta)$, $y=r \sin(\theta)$ and ${\cal L}_{\partial_{\theta}}$ denotes Lie differentiation with respect to the vector field $\partial_{\theta}$. Such a basis takes the form:
\begin{eqnarray*}
	\Pi^{\mathrm{CE},\pm1}		&=&	d(x\pm i y) \wedge d t,\\[0.2cm]
	\Pi^{\mathrm{CE},0}			&=&	d z \wedge d t,\\[0.2cm]
	\Pi^{\mathrm{CM},\chi }	&=& \star\, \Pi^{\mathrm{CE},\chi }
\end{eqnarray*}
The index $s$ indicates that the CE (CM) chiral family contain electric (magnetic) fields that are orthogonal to the $z-$axis when $\chi = 0$. Furthermore, the rationale for the $\chi-$labelling of solutions $F^{s,\chi}= d\(\star\,d(\alpha\Pi^{ s, \chi} )\)$  in this chiral basis follows from an exploration of the spacetime proper-time $\tau$ parameterised curves $x^{\mu}=C^{\mu}(\tau)$ of massive test particles with electric charge $q$ and mass $m_{0}$ interacting with such electromagnetic fields according to the covariant Lorentz force equations of motion:
\begin{eqnarray}
	\begin{split}\label{EOM}
		\nabla_{\dot{C}}\dot{C} &\;=\; \frac{q}{m_{0}c_{0}^{2}}\, \wt{i_{\dot{C}}F}\\[0.2cm]
		g(\dot{C},\dot{C}) &\;=\; -1
	\end{split}
\end{eqnarray}
where $\wt{i_{\dot{C}}F}=g^{-1}(  i_{\dot{C}}F  , -)$ and $i_{\dot{C}}$ denotes the interior contraction operator on forms, with respect to the tangent vector $\dot{C}$.

\LASERSPACECURVES{1}{%
	Three-dimensional space-curves for particles subject to an incident $(\text{CM},1)$ laser pulse (left), $(\text{CM},0)$ laser pulse (centre) and $(\text{CM},-1)$ laser pulse (right) with parameters $\{\Lambda=600, \Psi_{1}=1, \Psi_{2}=1000, \Phi=0.001, \Xi=1\}$. Each particle has initial velocity $\{{\dot{R}(0)=0}, {\dot{\theta}(0)=0}, {\dot{Z}(0)=1/200}\}$. The shaded  circular disc region indicates the initial  spot size ($R=10000$ for $(\text{CM},\pm 1)$ laser pulses and $R=20000$ for a $(\text{CM},0)$ laser pulse) relative to the black markers on the space-curves that denote the initial positions of the charged test particles.
	}
	
To facilitate this numerically one introduces  dimensionless variables. The Minkowski metric tensor field $g$ above has MKS physical dimensions $[L]^2$. The MKS dimension of electromagnetic quantities follows by assigning to $F$ in any coordinate system the dimension $[Q/\epsilon_{0}]$. Furthermore, in terms of Minkowski polar coordinates $\{ t,r,z,\theta \}$, introduce  the dimensionless coordinates $\{{R=r/\Phi\W}, {T=c_{0} t/\W}, Z=z/\Xi\W\}$ and dimensionless parameters $\Lambda, \Psi_{j}=\psi_{j}/\W$ ($j=1,2$) where ${[\Psi_{j}]= [\Phi]=[\Xi]=1},{[\W]=[L]}$. Then with the dimensionless complex scalar field
$\widehat{\alpha }(T,R,Z)=\alpha(t,r,z)$: 
\begin{eqnarray*}
	A &=& \frac{m_{0} c_{0}^{2} \W^{3}\,\Lambda}{q} \star d\left(\widehat{\alpha} \, \widehat{\Pi}\right) 
\end{eqnarray*}
for a choice of dimensionless covariantly constant tensor $\widehat{\Pi}$ so that $[A]=[Q/\epsilon_{0}]$. \\ 

A choice of dimensionless parameters can be used to solve (\ref{EOM})   numerically  for a collection of trajectories for charged particles, each arranged initially around the circumference of a circle in a plane orthogonal to the propagation axis of incident CM type electromagnetic pulses with different chirality. The resulting space-curves in 3-dimensions, displayed in figure 1, clearly exhibit the different responses of charged matter to CM pulses with distinct chirality values \cite{GTW_Lasers}.  Similar space-curves arise from charged particle interacting with chiral CE type modes .

\section{Gravitational Pulses in Vacua}

In any matter-free domain of spacetime $\mathcal{U}\subset\MAN{\wh{g}}$, an Einsteinian gravitational field is described by a real symmetric covariant rank-two tensor $\wh{g}$ with Lorentzian signature that satisfies the vacuum Einstein equation 
\begin{eqnarray*}
	\EIN{\wh{g}} &=& 0 
\end{eqnarray*}	
where
\begin{eqnarray}\label{VAC}
	\EIN{\wh{g}} &=& \RIC{\wh{g}} - \frac{1}{2}\,\TR{\wh{g}}(\RIC{\wh{g}})\,\wh{g}
\end{eqnarray}
and $\RIC{\wh{g}}$ is the Levi-Civita Ricci tensor associated with the torsion-free, metric-compatible Levi-Civita connection $\NABLA{\wh{g}}$. A coordinate independent linearisation of (\ref{VAC}) about an arbitrary Lorentzian metric  can be found in \cite{stewart,TuckerGEM}. In particular a linearisation about a flat Minkowski spacetime metric $\eta$ on $\mathcal{U}$ determines the linearised metric\footnote{%
	Physical dimensions of length${}^{2}$  are assigned to the {\it tensors} $g$ and $\eta$. The Ricci-scalar associated with $g$ then has the dimensions of length${^{-2}}$. In a $g$-orthonormal coframe the components of $g$ are $\{-1,1,1,1\}$ and in an $\eta$-orthonormal coframe the components of $\eta$ are $\{-1,1,1,1\}$. This does not imply that components of  the tensor field $\eta$ are necessarily constant in an arbitrary coframe on $\UU$.} %
$g=\eta + h$ and to first order one writes $\wh{g} = g + O(\kappa^2).$  The variable  $\kappa$ is a  dimensionless parameter in $h$ used to keep track of the expansion order and
\begin{eqnarray}\label{ETA}
	\eta &=& -\, e^{0} \tensor e^{0} + \sum_{k=1}^{3} e^{k} \tensor e^{k} \;=\; \eta_{ab}\,e^{a}\tensor e^{b}
\end{eqnarray}
in {\it any}  $\eta-$orthonormal coframe\footnote{%
	An arbitrary coframe is a set of $1-$ forms  $\{e^{a}\}$ satisfying $e^{0}\wedge e^{1}\wedge e^{2}\wedge e^{3} \neq 0 $.  If $\beta=\beta_{ab}e^{a}\tensor e^{b}$ in {\it any} coframe $\{e^{a}\}$ on $\mathcal{U}$, $\TR{\eta}(\beta)=\beta_{ab}\eta^{ab}$ with $\eta^{ab}\eta_{bc}=\delta^{a}_{c}$.%
	} 
on $\UU$. Since we explore the source-free Einstein equation (relevant to the motion of test-matter far from any sources) the scale associated with any linearised solutions must be fixed by the solutions themselves rather than any coupling to self-gravitating matter. Furthermore since only dimensionless relative scales have any significance we define the tensor $h$ to be a {\it perturbation} of $\eta$ on $\UU$ relative to {\it any} local $\eta-$orthonormal coframe $\{e^{a}\}$  provided 
\begin{eqnarray}\label{CNDS}
	|\,h(X_{a},X_{b})\,| \;<\; 1 \qquad \text{on \;$\mathcal{U}$} \qquad  \text{for all $a,b=0,1,2,3$}
\end{eqnarray} 
where  $e^{a}\in T^{*}\mathcal{U}$, $X_{b}\in T\mathcal{U}$, $e^{a}(X_{b})=\delta^{a}_{b}$. It should be noted that the perturbation order of any component of the $\eta$-covariant {\it derivative} of a tensor and its $\eta$-trace relative to such a coframe is not necessarily of the same order as that assigned to the tensor. Thus perturbation order is not synonymous with ``scale'' in this context. We use the conditions (\ref{CNDS}) to {\it define} perturbative Lorentzian spacetime to be sub-domains $\UUP\subset \UU$ where 
\begin{eqnarray*}
	\mathop{\text{max}}_{a,b} \,\vert h(X_{a},X_{b})\vert &<& 1.
\end{eqnarray*}
The real tensor $h\equiv\TRREV{\eta}(\psi')$ with $\psi'\equiv\mathrm{Re}(\psi)$ may be constructed from any complex covariant symmetric rank two tensor $\psi$ satisfying \cite{TuckerGEM}:
\begin{eqnarray}
	\LAP{\eta}(\psi) - 2\TRREV{\eta}\(\text{Sym}\NABLA{\eta}(\,\DIV{\eta}(\psi)\,) \) &=& 0.
\end{eqnarray}
Here and below, $\NABLA{\eta}$ denotes the operator of Levi-Civita covariant differentiation associated with $\eta$,  $X^{a}\equiv\eta^{ab}X_{b}$, $Y\equiv\NABLA{\eta}_{X_{a}}X^{a}$ and for all covariant symmetric rank two tensors  $T$ on $\UU$: 
\begin{eqnarray*}
	\LAP{\eta}(T)		&\equiv& \NABLA{\eta}_{Y}T - \NABLA{\eta}_{X_{a}}\NABLA{\eta}_{X^{a}}T	\\[0.2cm]
	\DIV{\eta}(T)		&\equiv& (\NABLA{\eta}_{X_{a}}T)(X^{a},-) \\[0.2cm]
	\TRREV{\eta}(T)		&\equiv& T - \frac{1}{2}\TR{\eta}(T)\,\eta . 
\end{eqnarray*}
Since for any $g$ the reverse-trace map $\TRREV{g}$ satisfies $\TRREV{g}\circ\TRREV{g}=\text{Id}$, if $\psi'$ is trace-free with respect to $\eta$, then $h=\psi'$. If $\psi$ is also divergence-free with respect to $\eta$, then $\LAP{\eta}(\psi)=0$. Thus, divergence-free, trace-free solutions $\psi$ satisfy:
\begin{eqnarray*}\label{DFTF_LAP_EQ}
	\LAP{\eta}(\psi) \;=\; 0 \qquad\text{and}\qquad	\DIV{\eta}(\psi) \;=\; 0.
\end{eqnarray*}
Given $\psi$ and hence $g=\eta + h$, all proper-time parametrised time-like spacetime geodesics $C$ on $\mathcal{U}$, with tangent vector $\dot{C}$, associated with $g$,  must satisfy the differential-algebraic system
\begin{eqnarray}
\begin{split}\label{DAE}
	\NABLA{g}_{\dot{C}}\dot{C} &\;=\; 0 \\[0.2cm]
	g(\dot{C},\dot{C}) &\;=\; -1.
\end{split}
\end{eqnarray}	 
If any  worldline $C$ has components $C^{\mu}(\tau)$ in any local chart on $\mathcal{U}$ with coordinates $\{x^{\mu}\}$  and  $ {\dot C}^{\mu}=\partial_{\tau} C^{\mu} $ then
\begin{eqnarray*}\label{ACCEL_EQ}
	\NABLA{g}_{\dot{C}}\dot{C} \;\;\equiv\;\; \frac{D\dot{C}^{\mu}}{d\tau} \,\partial_{\mu} &\;=\;& 
   \( \frac{d\dot{C}^{\mu}}{d\tau} +(\Gamma^{\mu}_{\alpha\beta} \circ C) \,\dot{C}^{\alpha}\dot{C}^{\beta}\)\partial_{\mu}
\end{eqnarray*}
where $\Gamma^{\mu}_{\alpha\beta}$ denotes a Christoffel symbol associated with $\NABLA{g}$.\\

In the following only solutions to (\ref{DAE}) that lie in the perturbative domains $\UUP$ are displayed. The worldline of an idealised {\it observer} in $\mathcal{U}$ is modelled by the integral curve $C_{V}$ of a future-pointing time-like unit vector field $V$, (i.e. $g(V,V)=-1$).  At any event in $\mathcal{U}$ the $g-$orthogonal decomposition of $\dot{C}$ with respect to an observer $C_{V}$:
\begin{eqnarray*}
	\dot{C} &=& {\boldsymbol \nu} - g(\dot{C},V) \, V,       
\end{eqnarray*}
with $g({\boldsymbol \nu},V)=0$ defines the Newtonian 3-velocity field $\mathbf{v}$ on $C$ relative to the integral curve  $C_{V}$  that it intersects in spacetime: 
\begin{eqnarray*}
	\mathbf{v} &=& \frac{\boldsymbol\nu}{g( \dot{C},V)} \;\;\equiv\;\; \frac{\dot{C}}{g( \dot{C},V)} + V. 
\end{eqnarray*}
Relative to $C_{V}$, the observed ``Newtonian speed''  of the proper-time parameterised worldline $C$  at any event is then $v\equiv\sqrt{g(\mathbf{v},\mathbf{v})\,}$. If $\NABLA{g}_{V}V=0$, the observer is said to be {\it geodesic} otherwise it will be accelerating. If there exists a local coordinate system $\{t,\xi_{1},\xi_{2},\xi_{3}\}$ on $\mathcal{U}$  with $\partial_{t}$  time-like and in which $C_{V}$ can be parameterised  monotonically with $\lambda$ as $t=\lambda, \xi_{1}=\xi_{1}(0), \xi_{2}=\xi_{2}(0),\xi_{3}=\xi_{3}(0)$ then such an observer is said to be at rest in this coordinate system. Although {\it any} particular time-like worldline defines a local ``rest observer'' in {\it some chart}, only the existence of a {\it family} of rest observers in a particular chart $\Phi_{\mathcal{U}}$ on  $\mathcal{U}$ provides a way to interpret the Newtonian velocity of any event on a time-like worldline that is not necessarily a rest-observer in $\mathcal{U}$. In units\footnote{%
   We introduce a length scale parameter $L_{0}$ to relate coordinates $\{t,r,z\}$ with physical dimensions $\{\text{time, length, length}\}$ respectively to the dimensionless variables $\{T,R,Z\}$:
		\begin{eqnarray*}
			r &=& L_{0}R, \qquad z \;=\; L_{0}Z, \qquad c_{0}t \;=\; L_{0}T 
		\end{eqnarray*}
	where $c_{0}$ is a fundamental constant with the physical dimensions of speed. In SI units $c_{0}=3\times 10^{8}$ m/sec. In this scheme the parameter $\tau$ has length dimensions and its conversion to a parameter $\tau^{\prime}$ with dimensions of a clock time is given by $\tau^{\prime}=\tau/c_{0}=\wh{\tau} L_{0}/c_{0}$  where $\wh{\tau}$ is a dimensionless parameter.} 
with $c_{0}=1$, a point particle of rest-mass $m_{0}$ with a worldline $C$, when observed by $C_{V}$,  has energy and 3-momentum  values at any event on $C$  given by  $\mathcal{E}_{V} = \gamma_{V} m_{0}$ and $\mathbf{p}_{V}=\gamma_{V} m_{0}\mathbf{v}$ respectively, where $\gamma_{V}\equiv 1/\sqrt{1- g( \mathbf{v}, \mathbf{v} )\,}$.\\

The properties of the scalar field $\alpha$ needed to generate chiral, wave-like and pulse-like solutions to the linearised source-free Einstein equations have been developed in \cite{TW_GPulses}. To emulate the methodology used above to uncover the electromagnetic pulse-like solutions we note that a key role is played by the commutativity of certain exterior operators with covariantly constant {\it antisymmetric} tensors and the Hodge de-Rham Laplacian $\delta d$ on scalar fields in Minkowski spacetime. In Einsteinian gravitation one seeks similar properties involving the tensor Laplacian $\LAP{g}$, $\nabla^{(g)}$ and {\it symmetric} tensors in spacetimes with a metric tensor $g$. The key general identity is the relation between $\LAP{g}$, $\nabla^{(g)} $ and the curvature operator $R^{(g)}_{X,Y}$ of the torsion-free, metric compatible connection $\nabla^{(g)} $:
\begin{eqnarray*}
	\LAP{g} \(\nabla^{(g)} \alpha\) &=&	\nabla^{(g)}\LAP{g}(\alpha) + R^{(g)}_{ X_{j}, \widetilde{d \alpha}} \,e^{j}
\end{eqnarray*}
where $\alpha$ is a scalar field. Therefore, on Minkowski spacetime with metric $\eta$, the commutation relation $[\LAP{\eta},\nabla^{(\eta)}]\alpha=0$ on scalar fields $\alpha$ and if $\alpha$ is {\it any}  complex (four times differentiable)  scalar field on $\mathcal{U}$ then 
\begin{eqnarray*}
	\NABLA{\eta} \NABLA{\eta}\,\LAP{\eta}(\alpha) \,=\, \LAP{\eta}\!\(\NABLA{\eta} \NABLA{\eta}\alpha\).
\end{eqnarray*}
Hence, since one may show that $\LAP{\eta}(\alpha)=-\delta d\alpha $, then $\LAP{\eta}(\psi)=0$ provided $\delta d \alpha = 0$. Furthermore, since
\begin{eqnarray*}
	\DIV{\eta}(\nabla^{(\eta)}\beta) \,=\, \LAP{\eta}(\beta) \qquad \text{for all $1-$forms $\beta$},
\end{eqnarray*}
and $\nabla^{(\eta)}\alpha = d\alpha$ for all $0-$forms $\alpha$, it follows that in Minkowski spacetime with the Levi-Civita connection $\nabla^{(\eta)}$ that $[\LAP{\eta},d]\alpha=0$ and with\footnote{%
	For any scalar $\alpha$  and metric tensor $g$, $\NABLA{g}\alpha=d\alpha$ is independent of $g$.%
	} 
$\psi=\NABLA{\eta}d\alpha$: 
\begin{eqnarray*}
	\DIV{\eta}(\psi)\,=\,\DIV{\eta}(\nabla^{(\eta)}d\alpha) \,=\, \LAP{\eta}(d\alpha) \,=\, d(\LAP{\eta}(\alpha)) \,=\, -d(\delta d\alpha),
\end{eqnarray*}
which vanishes when $\delta d\alpha=0$. Hence $\LAP{\eta}(\psi)=0$ and $\DIV{\eta}(\psi)=0$ are both satisfied for {\it any} suitably differentiable complex scalar field $\alpha$ satisfying $\delta d \alpha=0$. This should be compared with (\ref{BOX}), the equation determining the class of electromagnetic field solutions discussed above. 
Furthermore since $\NABLA{\eta}$ is the torsion-free Levi-Civita connection, $\psi$ is symmetric and trace-free with respect to $\eta$. Hence, such complex $\alpha$ in general give rise to real metrics: 
\begin{eqnarray*}
	g &\;=\;& \eta + \psi'.
\end{eqnarray*}	
In a local chart $\Phi_{\mathcal{U}}$ possessing dimensionless coordinates $\{T,R,\theta,Z\}$ with $T\geq 0$, $R>0$,
$\theta\in[0,2\pi)$ and $|Z|\geq 0 $ on a spacetime domain $\mathcal{U}\subset \mathcal{M}^{(\wh{g})}$, a local coframe $\CF$ adapted to these co-ordinates is $\{e^{0}=dT,\, e^{1}=dR, \,e^{2}=R\,d\theta, \,e^{3}=dZ\}$.  With $\eta$ given by (\ref{ETA}), this coframe is $\eta-$orthonormal but not in general $g-$orthonormal. Such a chart facilitates the coordination of a series of massive test particles initially arranged in a series of concentric rings with different values of $R$ lying in spatial planes with different values of $Z$ at $T=0$. Furthermore, we define for any  metric $g=\eta+h$ on $\UU$:
\begin{eqnarray*}
	\HMAX(T,R,\theta,Z) &\equiv& \mathop{\text{max}}_{0\leq a,b\leq 3} \,\left\vert \frac{}{}h(X_{a},X_{b})\right\vert.
\end{eqnarray*}
The particular complex scalar $\alpha$ of relevance here satisfying $\LAP{\eta}(\alpha)=0$ is given in the $(T,R,\theta,Z)$ chart $\Phi_{\UU}$ above as\footnote{%
	This should be compared with (\ref{alp}), the scalar field for the pulsed Maxwell solutions which has the same form.}
\begin{eqnarray*}
	\alpha(T,R,Z) &=& \frac{\kappa}{R^{2}+Q_{12}(T,Z)} 
\end{eqnarray*}
where 
\begin{eqnarray*}
	Q_{12}(T,Z) \;\equiv\; \(\df Q_{1}+i(Z-T)\)\(\df Q_{2}-i(Z+T)\)
\end{eqnarray*}
and $\kappa$, $Q_{1}$, $Q_{2}$ are strictly positive real dimensionless constants. The scalar $\alpha(T,R,Z)$ is then singularity-free in $T$, $R$ and $Z$ and clearly axially-symmetric with respect to rotations about the $Z-$axis. It also gives rise to an axially-symmetric complex tensor $\psi_{0}$ satisfying\footnote{%
	Since $\NABLA{\eta}$ is a flat connection, if $K$ is an $\eta-$Killing vector then the operator $\NABLA{\eta}\mathcal{L}_{K} = \mathcal{L}_{K}\NABLA{\eta}$  on all tensors.%
	} %
$\mathcal{L}_{\partial_{\theta}}\psi_{0}=0$. In $\mathcal{U}$, the real axially-symmetric metric tensor $g_{0}$ then has non-zero components in the coframe  $\mathcal{C}$:
\begin{eqnarray*}
	g_{00} &\;=\;& -1 + \partial^{2}_{TT}\,\alpha^{\prime} \\[0.1cm]
	g_{01} &\;=\;& g_{10} \;\;=\;\; \partial^{2}_{TR}\,\alpha^{\prime} \\[0.1cm]
	g_{03} &\;=\;& g_{30} \;\;=\;\; \partial^{2}_{TZ}\,\alpha^{\prime} \\[0.1cm]
	g_{11} &\;=\;& 1+\partial^{2}_{RR}\,\alpha^{\prime} \\[0.1cm]
	g_{13} &\;=\;& g_{31} \;\;=\;\; \partial^{2}_{RZ}\,\alpha^{\prime} \\[0.1cm]
	g_{22} &\;=\;& 1 + \partial_{R}\,\alpha^{\prime}\,/\,R \\[0.1cm]
	g_{33} &\;=\;& 1 + \partial^{2}_{ZZ}\,\alpha^{\prime}
\end{eqnarray*}
where $\alpha^{\prime}\equiv\mathrm{Re}(\alpha)$ and satisfies $\mathcal{L}_{\partial_{\theta}}g_{0}=0$.\\

Complex symmetric tensors $\psi_{m}$ with integer chirality $m>0$ satisfying $\LAP{\eta}(\psi_{m})=0$, $\DIV{\eta}(\psi_{m})=0$, $\TR{\eta}(\psi_{m})=0$ and $\frac{1}{i}\mathcal{L}_{\partial_{\theta}}\psi_{m}=m \psi_{m}$  may be generated from $\psi_{0}$ by repeated covariant differentiation with respect to a particular $\eta-$null  and $\eta-$Killing complex vector field  $S$:
\begin{eqnarray*}\label{DELDEL}
	\psi_{m} &=& \underbrace{\NABLA{\eta}_{S}\cdots\cdots\NABLA{\eta}_{S}}_{\text{$m$ times}}\psi_{0} 
\end{eqnarray*}
where 
\begin{eqnarray*}
	\eta(S,-) \;=\; d(\,Re^{i\theta}\,) \qquad\text{or explicitly:}\qquad S &\;=\;& e^{i\theta}\(\frac{\partial}{\partial R} + \frac{i}{R}\frac{\partial}{\partial\theta}\).
\end{eqnarray*}
Solutions with negative integer chirality can be obtained by complex conjugation of the positive chirality complex eigen-solutions. Each $\psi_{m}$ defines a real spacetime metric $g_{m}=\eta + \psi'_{m}$ on $\mathcal{U}$ which, for $m\neq 0$, is not axially symmetric:  $\mathcal{L}_{\partial_{\theta}}g_{m}\neq 0$.\\

\RRZP{0.9}{%
	An indication of the nature of the spacetime geometry determined by $g_{0}$ on $\mathcal{M}$  is given by the structure of the associated Ricci curvature scalar $\CVRR{0}$. Regions where $\CVRR{0}(T,1,Z)$ change sign are clearly visible in the right side where a 2-dimensional density plot shows a pair of prominent loci that separately approach the future ($T\geq 0$) light-cone of the event at $\{R=1,\,T=0,\,Z=0\}$. A more detailed graphical description of $\CVRR{0}(T,1,Z)$ is given in the  left hand 3-dimensional plot where an initial  pulse-like maximum around $T\simeq 0$ evolves into a pair of enhanced loci with peaks at values of $Z$ with {\it opposite signs} when $T\geq 1$. This Ricci curvature scalar is generated from a metric perturbation pulse with parameters $L_{0}= 1$, $Q_{1}=Q_{2}= 1$, $\kappa = 1/4$.%
	}

An indication of the nature of the spacetime geometry determined by $g_{m}$ on $\mathcal{M}^{(g_{m})}$  is given by the structure of the associated Ricci curvature scalar $\CVRR{m}(T,R,Z)$. Unlike gravitational wave spacetimes  this scalar is not identically zero. For $m=0$ it is axially symmetric and in the chart $\Phi_{\UU}$ its independence of  $\theta$  means that  for values of fixed radius $R_{0}$  its structure can be displayed for a range of  $T$ and $Z$  values given a choice of parameters $(Q_{1},Q_{2},\kappa,L_{0})$. Regions where $\CVRR{0}(T,1,Z)$ change sign are clearly visible in the right side of  figure~\ref{fig:rrzp} where a 2-dimensional density plot shows a pair of prominent loci that separately approach the future ($T\geq0$)  light-cone of the event at $\{R=1,T=0,Z=0\}$. A more detailed graphical description of $\CVRR{0}(T,1,Z)$ is given in the left hand 3-dimensional plot in figure~\ref{fig:rrzp} where an initial  pulse-like  maximum around $T\simeq 0$ evolves into a pair of enhanced loci with peaks at values of $Z$ with {\it opposite signs} when $T\geq 1$. In this presentation the maximum pulse height has been normalised to unity. This characteristic behaviour is similar to that possessed by $\mathrm{Re}(\,\alpha(T,1,Z)\,)$. It suggests that ``tidal forces'' (responsible for the geodesic deviation of neighbouring geodesics \cite{schutz,laemmerzahl,perlick}) are concentrated in spacetime regions where components of the Riemann tensor of $g_{0}$ have pulse-like behaviour in domains similar to those  possessed by $\CVRR{0}(T,R,Z)$.\\

Explicit formulae for $\CVRR{0}(T,R,Z)$ and $\HMAX(T,R,\theta,Z)$ are not particularly illuminating\footnote{%
	Since $g_{0}$ is axially-symmetric, the function $\HMAX$ is independent of $\theta$. 
	}. %
However, for fixed values of the parameters $(Q_{1},Q_{2},\kappa,L_{0})$, their values can be plotted numerically in order to gain some insight into their relative magnitudes in any perturbative domain $\mathcal{P}_{\mathcal{U}}$. With $Z$ fixed at zero, figure~\ref{fig:hmaxrrzp} displays such plots as functions of $R$ and a set of $T$ values. It is clear that in perturbative domains the curvature scalar may exceed unity. Since in general:
\begin{eqnarray*}
	\CVRR{0}(T,R,Z) &=& \mathcal{Q}(T,R,Z)\kappa^{2} + O(\kappa^{3})
\end{eqnarray*}	
where $\mathcal{Q}$ is a non-singular rational function of its arguments and the tensor $h$ is, by definition, of order $\kappa$, figure~\ref{fig:hmaxrrzp} demonstrates that relative tensor $\kappa-$orders are not, in general, indicators of their corresponding relative magnitudes. \\

\HMAXRRZP{0.8}{%
	The axially-symmetric expressions $\vert\CVRR{0}(T,R,0)\vert$ and $\HMAX(T,R,0,0)$ are plotted as functions of $R$ 
	for $T=0,\,0.25,\,0.5,\,0.75$ and parameters $L_{0}=1$, $Q_{1}=Q_{2}=1$, $\kappa=1/4$. Regions where the blue curves lie under the red dotted line denote perturbative regions $\mathcal{P}_{\mathcal{U}}$. The grey shaded regions clearly indicate curvature scalars that are greater in magnitude than unity despite lying within $\mathcal{P}_{\mathcal{U}}$ regions. 
	}

\ZPZLAYERS{0.8}{%
	On the left  six geodesics are shown emanating from  six locations with $\theta$ values $0, \pi/3, 2\pi/3, \pi, 4\pi/3, 5\pi/3  $  on a  ring  with radius $10^{-4}$ in the plane $Z=0.735$ and six from similarly arranged points on rings of the same radius at  $Z=0.245$, $Z=-0.245$ and $Z=-0.735$. The initial locations are not resolved in these figures. The 24 geodesics each evolve from $\tau=0$ to $\tau=10^{4}$ and clearly display an axially symmetric  bi-directional jet structure from the rings in conformity  with the expectations based on the spacetime structure of $\CVRR{0}(T,1,Z)$ in figure~\ref{fig:rrzp}. The figure on the right resolves the structure of this jet array for $0\leq\tau\leq 100$. All geodesics are generated with the additional initial  conditions $\RDO = 0$, $\ZDO = 0$, $\thDO = 0.4$ and the background perturbation pulse has parameters $L_{0}=1$, $Q_{1}=Q_{2}=1$, $\kappa = 1/6$.  A single uni-directional jet-array arises when only one ring is populated with matter. This figure demonstrates that the jets from the sources at $Z=\pm 0.245$ have a dimensionless aspect ratio $\mathcal{A}(10^{4})=64.7$ much greater than those produced from the sources at $Z=\pm 0.735$ where $\mathcal{A}(10^{4})=3.22$. %
	}

By modelling thick accretion disks by a finite number of massive point particles  occupying  a number of planar rings   the system (\ref{DAE}) has been explored numerically. Sets of  space-curves in space-like sections of the perturbative spacetimes defined above are displayed in the following figures for various choices of the dimensionless parameters  $L_{0}$, $Q_{1}=Q_{2}$, $\kappa$ and evolution proper-time. The resulting jet-like structures of these space-curves with maximal proper-time parameter $\taumax$ are quantified in terms of aspect ratios defined by:
\begin{eqnarray*}
	\AR &\;\equiv\;& \left| \frac{\wh{Z}(\taumax)-\wh{Z}(0)}{\wh{R}(\taumax)-\wh{R}(0)} \right|.
\end{eqnarray*}

\ZPZLAYERSQ{0.8}{%
	On the left, six geodesics are shown emanating from  six locations with $\theta$ values $0, \pi/3, 2\pi/3, \pi, 4\pi/3, 5\pi/3  $  on a  ring  with radius $10^{-4}$ in the plane $Z=0.735$ and six from similarly arranged points on rings of the same radius at $Z=-0.735$. The initial locations are not resolved in these figures. The 12 geodesics each evolve from $\tau=0$ to $\tau=10^{4}$ and clearly display an axially symmetric uni-directional jet structure from the rings. All geodesics are generated with the additional initial  conditions $\RDO = 0$, $\ZDO = 0$, $\thDO = 0.4$ and the background perturbation pulse has parameters $L_{0}=1$, $Q_{1}=1$, $Q_{2}=3$, $\kappa = 1/6$. The figure in the centre shows an oppositely directed jet evolving from similar initial conditions, but with initial $Z=0.245$ and $Z=-0.245$. The figure on the right displays the ``asymmetric'' jet structure obtained by merging both pairs of sources with {\it dominant} component belonging to the jet in the left-hand figure having aspect ratio $\mathcal{A}(10^{4})=178.8$.%
	}

\section{Concluding Remarks}

We have developed a class of electromagnetic pulse-like solutions to the source-free vacuum Maxwell equations and shown that they may be classified in terms of {\it chiral} eigenstates by bringing them into interaction with electrically charged systems composed of point particles. A similar method of analytically constructing gravitational pulse-like solutions of the source-free linearised Einstein equations with definite chirality has also been developed. We have then explored numerically the nature of the time-like geodesics in certain perturbative spacetime domains associated with a family of zero chirality gravitational pulse-like solutions. \\

Using suitably arranged massive test particles to emulate a thick accretion disc, together with a particular family of fiducial observers, we have displayed a number of characteristic features of these geodesics in such background metrics. Within the context of a non-dimensional scheme, solution parameters can be chosen that result in characteristic spatial jet-like patterns in three-dimensions. These have specific dimensionless aspect ratios relative to well-defined directions in a  background gravitational pulse and the corresponding orthogonal subspace.\\

For each zero chirality gravitational pulse incident at $T=0$ on a  bounded region of matter in the vicinity of the spatial plane $Z=0$ in three-dimensions, one finds that (with $Q_{1}=Q_{2}$) a {\it pair} of oppositely directed jet-like structures arise: i.e. a pair of time-like geodesic families with Newtonian speeds approaching  terminal values less than the speed of light for {\it both} $Z>0$ {\it and} $Z<0$. For a pulse with $Q_{1}\neq Q_{2}$, we have demonstrated the existence of a pair of uni-directional jet-like structures from particular initial conditions. In all these cases, the structures have well-defined aspect ratios that can be calculated numerically. The propagation characteristics for $T>0$ of the pulse responsible for these jet structures in space is discernible from features of the non-zero Ricci scalar curvature associated with the perturbed spacetime domains. \\

We have also stressed that by linearising only the gravitational field equations and analysing the {\it exact} geodesic equations of motion in perturbative spacetime domains, one can capture the full effects of ``tidal accelerations'' on matter produced by the curvature tensor (and its contractions) associated with the metric perturbations. This opens up the possibility of a gravito-ionisation process whereby extended electrically neutral micro-matter can be split into electrically charged components by purely gravitational forces, leading to modifications of matter worldlines by the presence of Lorentz forces. \\

We conclude that background spacetime metrics derived from complex chiral solutions of the linearised source free Einstein equations, separately or in superposition, may offer a non-Newtonian gravitational mechanism for the initialisation of a dynamic process leading to astrophysical jet structures emanating from compact matter distributions, particularly since it is unlikely that such phenomena originate from a unique set of initial conditions. \\

\section*{Acknowledgements}
RWT is grateful to the University of Bolton and the Cockcroft Institute for hospitality and to STFC (ST/G008248/1) and EPSRC (EP/J018171/1) for support. Both authors are grateful to V. Perlick for his comments. All numerical calculations have been performed using Maple 2015 on a laptop.


\end{document}